\documentclass[prb,twocolumn,floatfix]{revtex4}
\usepackage{graphicx}

\begin{document}
\title{%
%Physics of the Anderson-Kondo model
Relevance of quantum fluctuations in the Anderson-Kondo model
}
\author{Robert Peters}
\affiliation{Institute for Theoretical Physics, University of G\"ottingen, 
Friedrich-Hund-Platz 1,
37077 G\"ottingen, Germany}
\author{Thomas Pruschke}
\affiliation{Institute for Theoretical Physics, University of G\"ottingen, 
Friedrich-Hund-Platz 1,
37077 G\"ottingen, Germany}

\begin{abstract}
  We study a localized spin coupled to an Anderson impurity to model the
situation found in higher transition metal or rare earth compounds like e.g.\
LaMnO$_3$ or Gd monopnictides. We find that, even for large quantum numbers of the
localized spin, quantum fluctuations play an essential role for the
case of ferromagnetic coupling between the spin and the impurity
levels. For antiferromagnetic coupling, a description in terms of
a classical spin is appropriate.
\end{abstract}
\maketitle              

\section{Introduction}
Transition metal oxides show a fascinating complex behavior in their
electronic properties.\cite{imada:98} This complexity stems from the
interplay between the formation of narrow $3d$-bands leading to a
delocalization of these states on the one hand and the local part of the
Coulomb interaction between the $3d$-electrons tending to localize
them.\cite{imada:98}  While compounds of the early $3d$ elements like
e.g.\ LaTiO$_3$, which typically accommodate one $3d$ electron, can at
least qualitatively be described in
terms of a one-band Hubbard
model\cite{hubbard:63,gutzwiller:63,kanamori:63},
materials involving higher transition metal elements like LaMnO$_3$
or TlSr$_2$CoO$_5$
require the use of a model including the full $3d$ shell. 
In particular, to understand the magnetic properties and the
frequently occurring metal-insulator transitions\cite{imada:98} one has
to take into account the interplay between density- (``Hubbard $U$'')
and exchange-type (``Hund's $J$'')
contributions to the local Coulomb interaction. Note that similar
features can also be found in compounds involving higher rare earth
elements, for example the rare earth pnictides.

A particularly interesting example is La$_{1-x}$Ca$_x$MnO$_3$. Besides is complicated
phase diagram comprising a large variety of paramagnetic and
magnetically ordered metallic and insulating phases one finds a
colossal magneto-resistance (CMR).\cite{ramirez:97} In this cubic
perovskite the five-fold degenerate $3d$ level is split by crystal field into
three-fold degenerate $t_{2g}$, which have the lower energy, and
two-fold degenerate $e_g$ states. These states have to be filled with
$4-x$ electrons, nominally yielding a metal even for $x=0$. However,
taking into account the local Coulomb interaction, three of these
electrons will occupy the $t_{2g}$-states forming an $S=3/2$ high-spin
state due to
Hund's coupling, which interacts ferromagnetically with the electron
occupying the $e_g$ states.
Ignoring the Coulomb repulsion among the electrons in the $e_g$
subsystem, one encounters the well-known double-exchange
model\cite{zener:51}, which has been extensively studied as suitable
model for manganites (see e.g.\ references in Ref.\
\onlinecite{held:00c}). In most of these investigations, however, the
$t_{2g}$-spin was approximated by a classical moment to allow the use
of standard techniques like e.g.\ quantum
Monte-Carlo\cite{yunoki:98a,yunoki:98b,calderon:98,arovas:98,muha:99,motome:03a,motome:03b} (QMC).
Without such a replacement, one is typically restricted to low-order
diagrammatic techniques (Ref.\ \onlinecite{hickel:04} and references therein).

A more realistic treatment should of course also include the local Coulomb interaction
within the $e_g$ subsystem. Such a model has been proposed
recently\cite{held:00c} and studied in the framework of the dynamical
mean-field theory\cite{georges:96} (DMFT). Again, the $t_{2g}$-spin had to be
replaced by a classical moment to allow the solution of the DMFT
equations with QMC. 

In this paper we want to study the validity of approximating the
quantum spin by a classical object. To this end, we investigate the
simplest possible model, viz a quantum impurity model consisting of a
local orbital with interacting charge degrees of freedom coupled to a
non-interacting host and a spin.

While such a model surely cannot
access every aspect of the physics of the corresponding lattice model,
it is the basic ingredient in a DMFT calculation and thus
understanding its fundamental properties is of importance to properly interpret
results obtained in a DMFT calculation. Moreover, although such
a calculation will focus on local dynamics only, one can obtain at
least qualitative results about possible ordered phases,
too.\cite{georges:96} To this end it is viable to obtain a feeling how
the additional spin will modify local charge and spin properties.

We employ Wilson's numerical
renormalization group\cite{wilson:75,krish:80a} (NRG) to solve this
model. This technique allows to treat the model in the whole parameter
regime and in particular identify small energy scales if present.

The paper is organized as follows. In the next section we present the
model and briefly review its properties for a classical spin in the
limit of vanishing Coulomb interaction. The presentation of our
results follows in section III. The paper closes with a summary and
discussion.

\section{The Model}
The simplest model that allows to obtain an idea how the coupling to
an additional local spin-degree of freedom influences the properties
of correlated electrons is 
\begin{equation}\label{eq:model}
H=
\begin{array}[t]{l}
\displaystyle
\sum\limits_{k\sigma}\epsilon_{k}
c^\dagger_{k\sigma}c^{\phantom{\dagger}}_{k\sigma}
+\sum\limits_\sigma
\left(\epsilon_d+\frac{U}{2}d^\dagger_{-\sigma}d^{\phantom{\dagger}}_{-\sigma}\right)
d^\dagger_{\sigma}d^{\phantom{\dagger}}_{\sigma}\\[5mm]
\displaystyle
+\frac{V}{\sqrt{N}}\sum\limits_{k\sigma}\left(
c^\dagger_{k\sigma}d^{\phantom{\dagger}}_{k\sigma}+\mbox{h.c.}\right)\\[5mm]
\displaystyle
-J_K\sum\limits_{\alpha,\beta}
\vec{S}\cdot\vec{s}_d\;\;.
\end{array}
\end{equation}
The first three terms represent a conventional single-impurity
Anderson model,\cite{anderson:61,hewson:book} while the last
contribution introduces an additional spin degree of freedom which
couples to the local states via an exchange interaction. Similar
models have been studied with various techniques in connection with double quantum
dots.\cite{kim:01,kang:01,apel:04,lara:DQD,cornaglia:05,bonca:DQD}
However, in these cases the additional impurity was represented by a
correlated charge degree of freedom coupled via a hopping. In the
limit of half-filling and weak interimpurity hopping, this system maps
to our model with $S=1/2$ and vanishing antiferromagnetic $J_K$. This limit shows
interesting physics on its own,\cite{cornaglia:05,vojta:02} for
example two-stage Kondo screening or quantum phase
transitions. However, these models neither do allow for {\em
  ferromagnetic} $J_K$, large and possibly anisotropic $J_K$ nor spins
$S>1/2$, which are the particular cases we are interested in here.

Of course our model (\ref{eq:model}) does not fully represent the
situation found in e.g.\ LaMnO$_3$ since it lacks the orbital degrees
of freedom.  On the other hand, NRG calculations for
multi-orbital models are extremely expensive\cite{pruschke:05} and we
believe that as far as the qualitative aspects are concerned this simplification will not
substantially modify the validity of our observations for the more
complicated model.

For a classical spin $\vec S$, the model (\ref{eq:model}) can be
solved exactly for $U=0$.\cite{furukawa:94,furukawa:98} The result for the
single-particle Green function of the $d$ states is
\begin{equation}\label{eq:gflarges}
G_d(z)=\frac{1}{2}\left(
\frac{1}{z-\epsilon_d+i\Delta_0+J_K}+
\frac{1}{z-\epsilon_d+i\Delta_0-J_K}\right)
\end{equation}
where $\Delta_0=\pi N_FV^2$ and we assumed $|\vec S|^2=1$ and a flat
conduction density of states (DOS) of infinite
width and value $N_F$. The resultant DOS
$\rho_d(\omega)=-\frac{1}{\pi}\Im m G_d(\omega+i\delta)$ shows two
peaks of width $\Delta_0$ centered at $\epsilon_d\pm J_K$. Note
that one can view this result as spin-averaged DOS of
the SIAM at $U=0$ in a magnetic field $J_K$ and that
the result is independent of the sign of $J_K$. 

\section{Results}
\subsection{Classical limit\label{subsec:classical}}
Neither for a quantum spin nor for a classical spin and $U>0$ an exact
solution exists. However, it is tempting to extend the above
interpretation for a classical spin in the following way: Solve a standard SIAM in
a magnetic field of $J_K$ and average over the resulting spectra for
spin up and down. Note that this procedure again leads to results that
do not depend on the sign of $J_K$.

Since the additional spin enters only on the local level, we use the
standard NRG algorithm\cite{krish:80a,sakai:89,bulla:98b} to solve the impurity problem
and calculate physical quantities. To obtain reliable spectra in a
magnetic field, we furthermore employ the technique proposed by
Hofstetter.\cite{hofstetter:00} Since a discretization of the energy
axis introduced in the NRG leads to discrete spectra, a broadening
must be introduced to obtain smooth results for dynamical
quantities. Again, we follow the standard procedure here.\cite{sakai:89}

\begin{figure}[htb]
\begin{center}
\includegraphics[width=0.45\textwidth,clip]{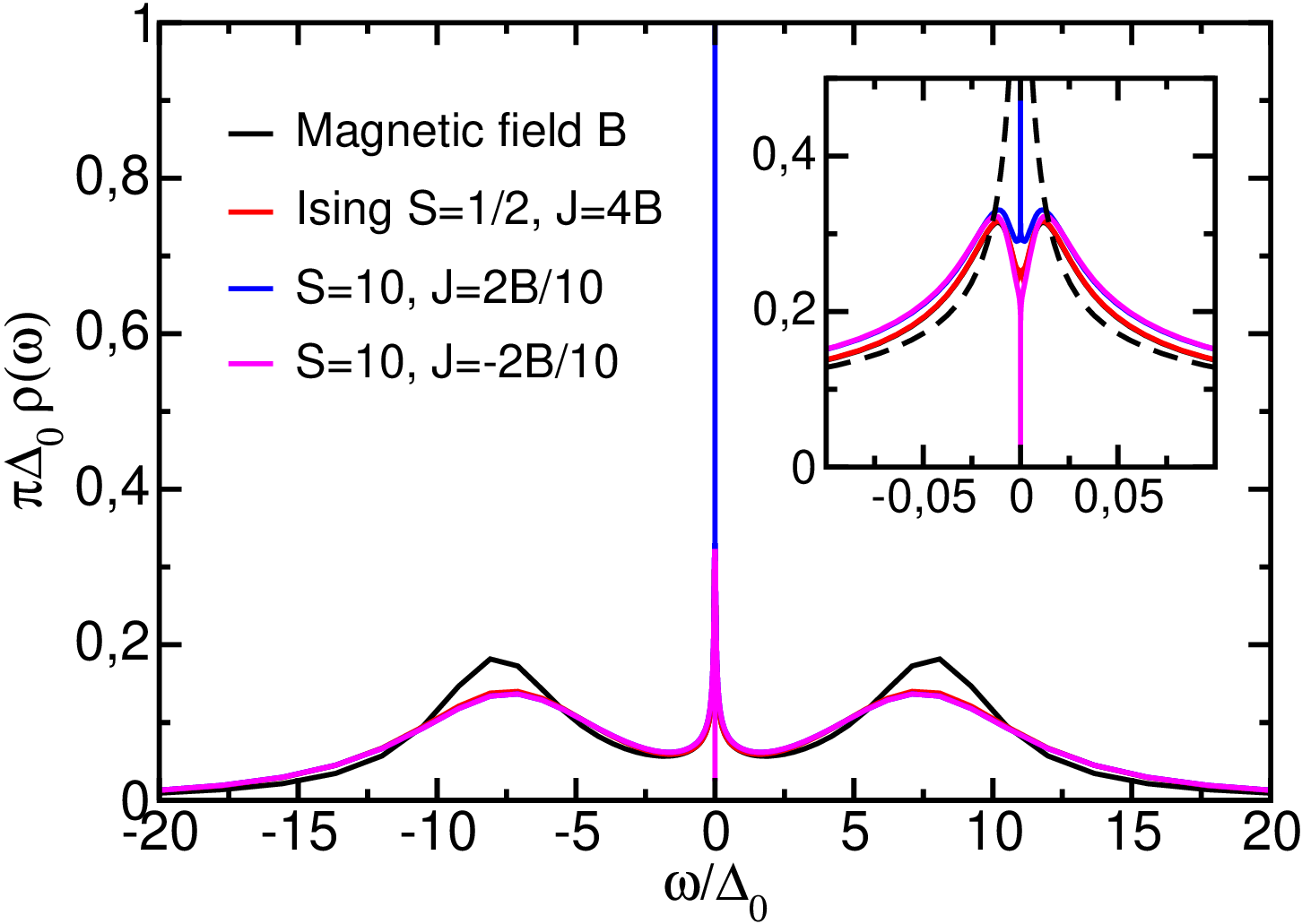}
\end{center}
\caption[]{Impurity DOS for SIAM in a magnetic field, with Ising spin
  $S=1/2$ and ``classical'' spin $S=10$. For the latter results for both
  antiferromagnetic and ferromagnetic coupling are shown. The inset
  displays an enlarged view of the region around $\omega=0$. The
  dashed black curve is the result for $S=0$ and $B=0$.\label{fig:compising}}
\end{figure}

As example we present in Fig.~\ref{fig:compising} calculations for the
SIAM
in a magnetic field, averaged over the spin direction, together with results for the SIAM with additional
Ising respectively classical spin. The parameters for the SIAM are $U/\pi\Delta_0=5.3$ and 
$\epsilon_d=-U/2$, i.e. particle-hole symmetry. The NRG
discretization parameter was $\Lambda=2.5$ and we kept
$1000\ldots4000$ states per iteration depending on the size of the
local spin. The NRG-spectra finally were broadened with a width
$b=0.6$.

The black curve in
Fig.~\ref{fig:compising} was obtained from a calculation with a magnetic
field $B=8\cdot10^{-3}\Delta_0$, the red curve with a local Ising spin
$S=1/2$ and coupling $J_K=4B$, the blue and magenta curves with local
spin $S=10$ and couplings $J_K=\pm 2B/10$ to simulate the
classical limit $S\to\infty$. The values of $J_K$ were scaled such
that $B=J_K s_dS$. The inset shows an enlarged view of the region
around $\omega=0$; for comparison, the dashed black curve represents the spectrum for
$S=0$, $B=0$. 

While the position of the Hubbard bands at higher energies coincides
for all curves, the distribution of spectral weight comes out different
for $S>0$ compared to the case with magnetic field and $S=0$. 
This deviation is a purely numerical effect related to the the differences in the distribution of
spectral weight in the Hubbard bands for the discrete NRG spectra with and without applied magnetic
field and the broadening introduced to obtain smooth spectra. 

In the
region around $\omega=0$ the calculation with Ising spin and
magnetic field coincide perfectly, yielding a splitting of the
original Kondo resonance at $B=J_K=0$, as expected. However, the
``classical limit'' with $S=10$  differs considerably. Depending on
the sign of $J_K$, either a Kondo resonance (ferromagnetic
coupling) or a gap (antiferromagnetic
case) emerges at $\omega=0$ in addition to the splitting of the
original Kondo peak. Apparently, even for such a large value of $S$ the effect of
quantum fluctuations is still prominent. However, the corresponding
energy scales are considerably reduced compared to $J_K=0$ and we
expect the results to converge to the anticipated one as $S\to\infty$.

\subsection{Quantum spins: $S=1/2$}
Let us now turn to the discussion of the effects of a local quantum
spin on the low-energy properties.
\begin{figure}[htb]
\begin{center}
\includegraphics[width=0.45\textwidth,clip]{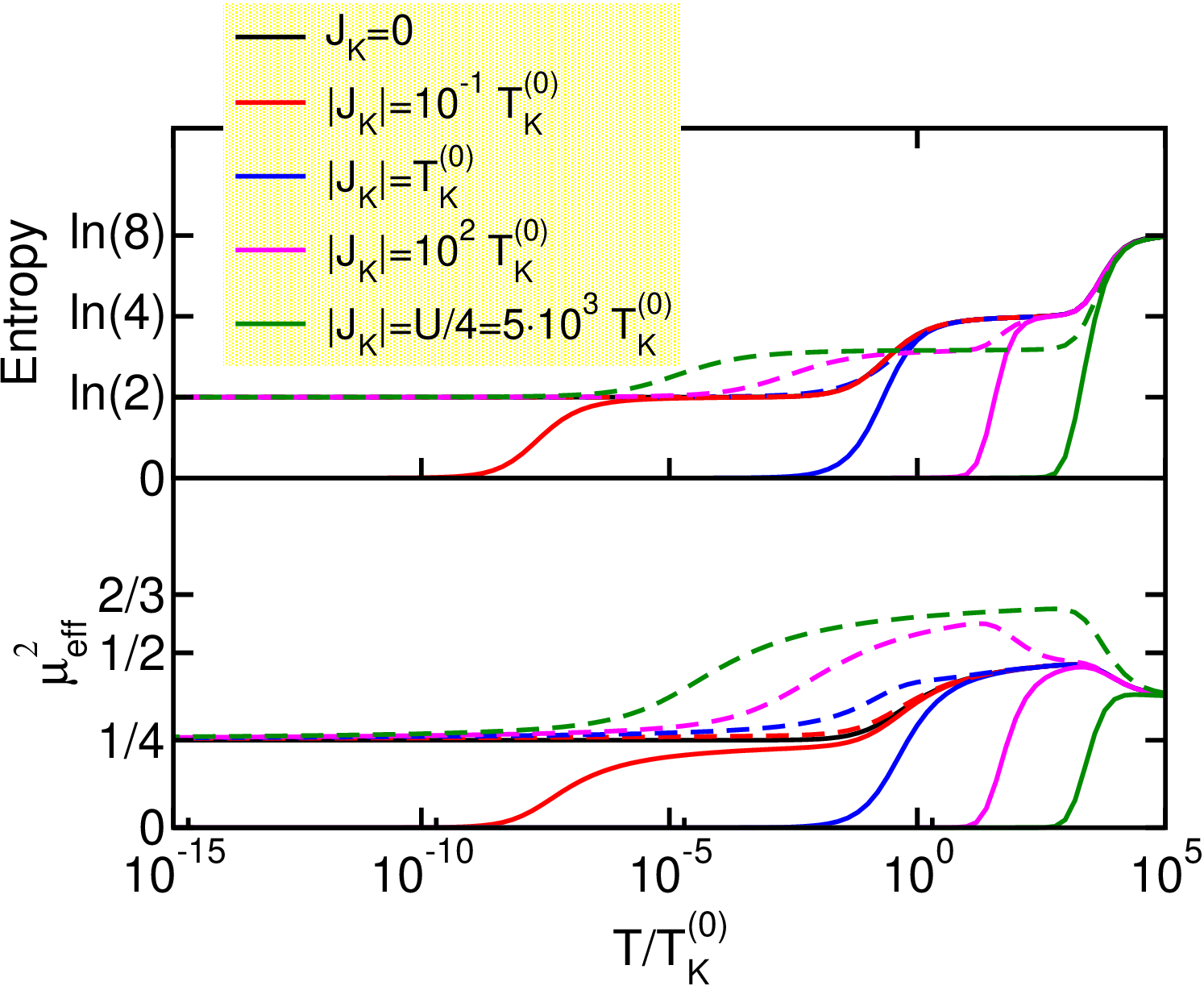}
\end{center}
\caption[]{Entropy and effective impurity moment for different values
  of $J_{\rm K}$ as function of $T/T^{(0)}_{\rm K}$. Full (dashed) lines
  denote antiferromagnetic (ferromagnetic) coupling.\label{fig:two_stage}}
\end{figure}
In Fig.~\ref{fig:two_stage} the impurity contribution to the entropy
(upper panel) and the effective impurity moment $\mu_{\rm eff}^2:=T\cdot\chi_{\rm
  imp}$ (lower panel) for different values of $J_K$ for a local spin
$S=1/2$ are shown as function of $T/T^{(0)}_{\rm K}$, where
$T^{(0)}_{\rm K}$ denotes the Kondo scale for
the system with $S=0$. The SIAM parameters were $U=6.4\pi\Delta_0$ at
particle-hole symmetry. The NRG discretization, number of states kept
etc.\ were chosen as before. The full lines in
Fig.~\ref{fig:two_stage} represent results for antiferromagnetic
coupling $J_{\rm K}$, the dashed lines those for ferromagnetic
coupling. 

Decreasing $J_{\rm K}$ from $J_{\rm K}=0$ to some antiferromagnetic $|J_{\rm K}|< T_{\rm
  K}^{(0)}$ results in the scenario depicted by the full red curves in
Fig.~\ref{fig:two_stage}. Around the temperature $T^{(0)}_{\rm K}$
screening occurs as in the normal Kondo effect, resulting in a
situation that resembles a free spin $S=1/2$ again. For a much lower 
$T_{\rm K}\propto T^{(0)}_{\rm K}\exp\left(-\alpha/(|J_{\rm
    K}|/T^{(0)}_{\rm K})\right)$, a second screening takes place to the
ground state with $S=0$. This latter can be viewed as a conventional
Kondo screening of the local spin by the fully formed local Fermi
liquid. Hence, the factors $T^{(0)}_{\rm K}$ appearing in the formula
for $T_{\rm K}$, representing the effective bandwidth respectively
DOS at the Fermi level of the quasi-particles of the local Fermi
liquid playing the part of the ``conduction states''. This effect has been observed before by several
authors\cite{vojta:02,cornaglia:05,bonca:DQD} and baptized two-stage
Kondo screening.
When $J_{\rm K}$ becomes of the order of $T^{(0)}_{\rm K}$, the Kondo
screening is replaced by the formation of a local singlet with an
energy scale $\approx |J_{\rm K}|$. 

For ferromagnetic coupling $J_{\rm K}>0$, on the other hand, the
proper fixed point is -- as in the conventional Kondo model -- the
local-moment one with residual entropy $\ln(2)$ and effective moment
$1/2$. Again, from the dashed curves in Fig.~\ref{fig:two_stage} one can distinguish two regimes. For $J_{\rm K}< T_{\rm
  K}^{(0)}$ we observe screening on the scale of $T_{\rm K}^{(0)}$,
the additional local spin effectively behaving like a free spin all
the way down to $T=0$. However, for $J_{\rm K}\gg T_{\rm K}^{(0)}$ the
impurity first forms a local spin triplet (entropy $\ln(3)$ and moment
$2/3$). This moment is then partially screened at a reduced Kondo
scale $T_{\rm K}\ll T_{\rm  K}^{(0)}$; energy scales and physical
properties will behave as in the conventional underscreened Kondo
model.\cite{nozieres:80,georges:97}

\begin{figure}[htb]
\begin{center}
\includegraphics[width=0.45\textwidth,clip]{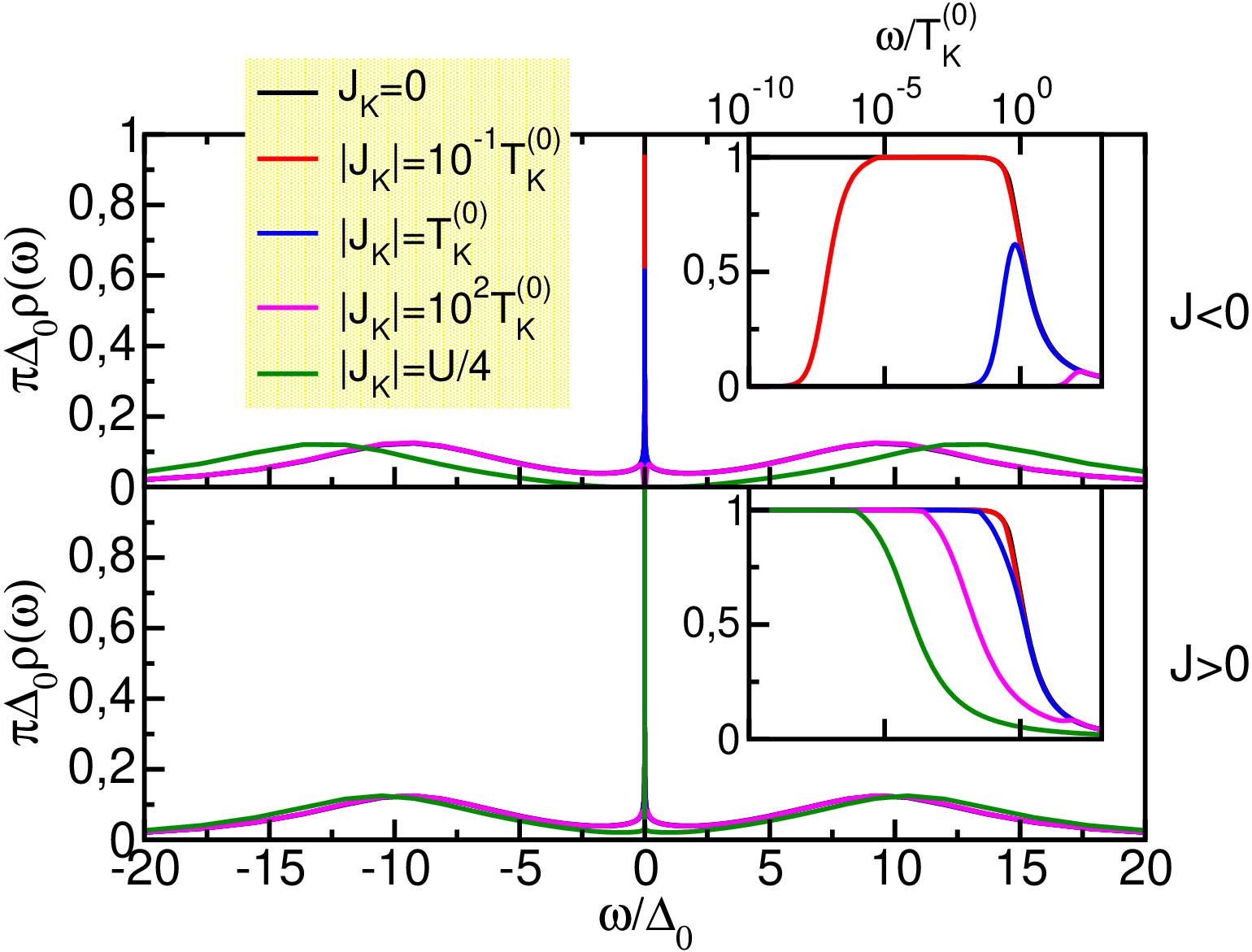}
\end{center}
\caption[]{Local DOS for different values
  of $J_{\rm K}$ as function of $\omega/\Delta_0$. The insets show
the spectra for $\omega>0$ in a semi-logarithmic plot.\label{fig:dos_half}}
\end{figure}
The behavior discussed previously is reflected in the local density
of states (DOS) depicted in Fig.~\ref{fig:dos_half}. As already noted
in the classical limit, the coupling to the additional spin leads to a
corresponding shift of the upper Hubbard band, which however is for larger
$|J_{\rm K}|$ more pronounced for antiferromagnetic exchange. In this
case on also nicely sees the two-stage screening
at $|J_{\rm K}|<T_{\rm K}^{(0)}$ and the formation of the local
singlet at $|J_{\rm K}|>T_{\rm K}^{(0)}$ (inset to upper panel of Fig.~\ref{fig:dos_half}), suppressing the Kondo
screening. Here, one always finds a gap in the DOS at $\omega=0$,
which size is set by $T_{\rm K}$.
For ferromagnetic coupling, on the other hand, the inset in the lower
panel of Fig.~\ref{fig:dos_half} proves that the screening resonance
remains intact, but shows a width decreasing with increasing $J_{\rm K}$.

\subsection{Quantum spins: General $S$}
\begin{figure}[htb]
\begin{center}
\includegraphics[width=0.45\textwidth,clip]{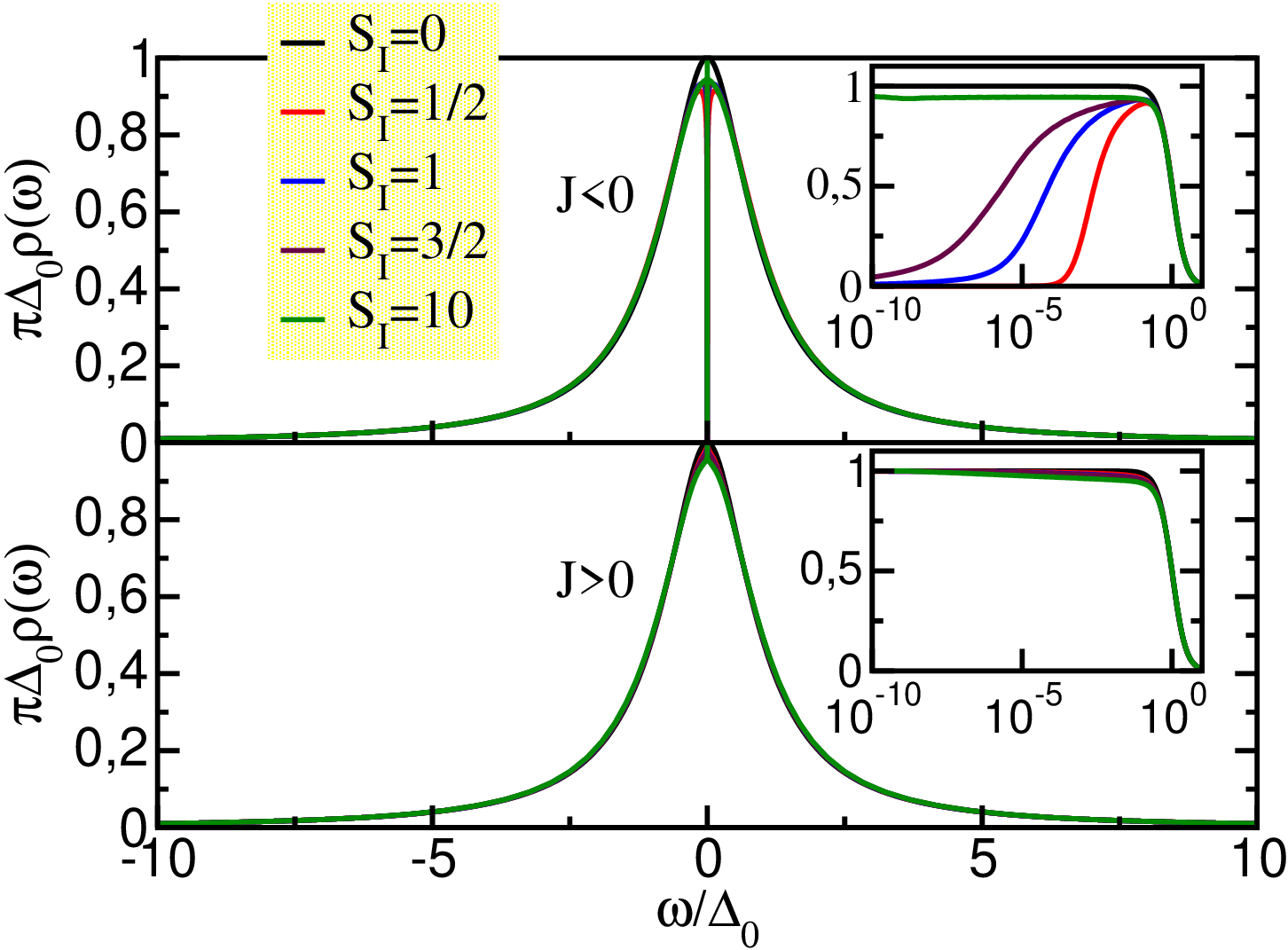}
\end{center}
\caption[]{Local DOS for fixed $\sqrt{S_I(S_I+1)}|J_{\rm K}|/\Delta_0=1/3$ 
as function of $\omega/\Delta_0$ in the weak-coupling regime
$U/\Delta_0=1$ for different values of local spin
$S_I$. The upper panel collects the results for antiferromagnetic
coupling, the lower panel those for ferromagnetic.
The insets show
the spectra for $\omega>0$ in a semi-logarithmic plot.\label{fig:dos_weak_c}}
\end{figure}

How does the behavior discussed in the previous section change with
increasing
spin quantum number $S_I$ or more precisely, do we recover a ``classical''
result for large enough $S_I$?
Let us start with a discussion of the weak-coupling results shown in Fig.~\ref{fig:dos_weak_c}.
The calculations were done in the weak-coupling regime $U/\Delta_0=1$ for fixed value
$\sqrt{S_I(S_I+1)}|J_{\rm K}|/\Delta_0=1/3$ to achieve the same
classical energy scale for all $S_I$. We did calculations up to
$S_I=10$, which, according to our results in section
\ref{subsec:classical}, we expect to be already very close to the
classical limit. 
Indeed, for $S_I=10$ (green curves in Fig.~\ref{fig:dos_weak_c}) we do
find almost identical behavior for $J_{\rm K}<0$ and $J_{\rm K}>0$ except for
extremely low temperatures. Note however, that even for this large
value for $S_I$ the DOS for $\omega/\Delta_0<0.1$ for $J_{\rm K}<0$
does not reach the full unitary limit due to quantum fluctuations.

The differences are more dramatic for small values of $S_I$.  As expected, for
$J_{\rm K}<0$ there occurs a Kondo screening with an energy-scale
$T_{\rm K}(S_I)$ decreasing exponentially with increasing $S_I$. Note that even
for comparatively large $S_I=3/2$ the influence of quantum
fluctuations is still pronounced and appears in an energy
regime that may still be of experimental relevance. 

\begin{figure}[htb]
\begin{center}
\includegraphics[width=0.45\textwidth,clip]{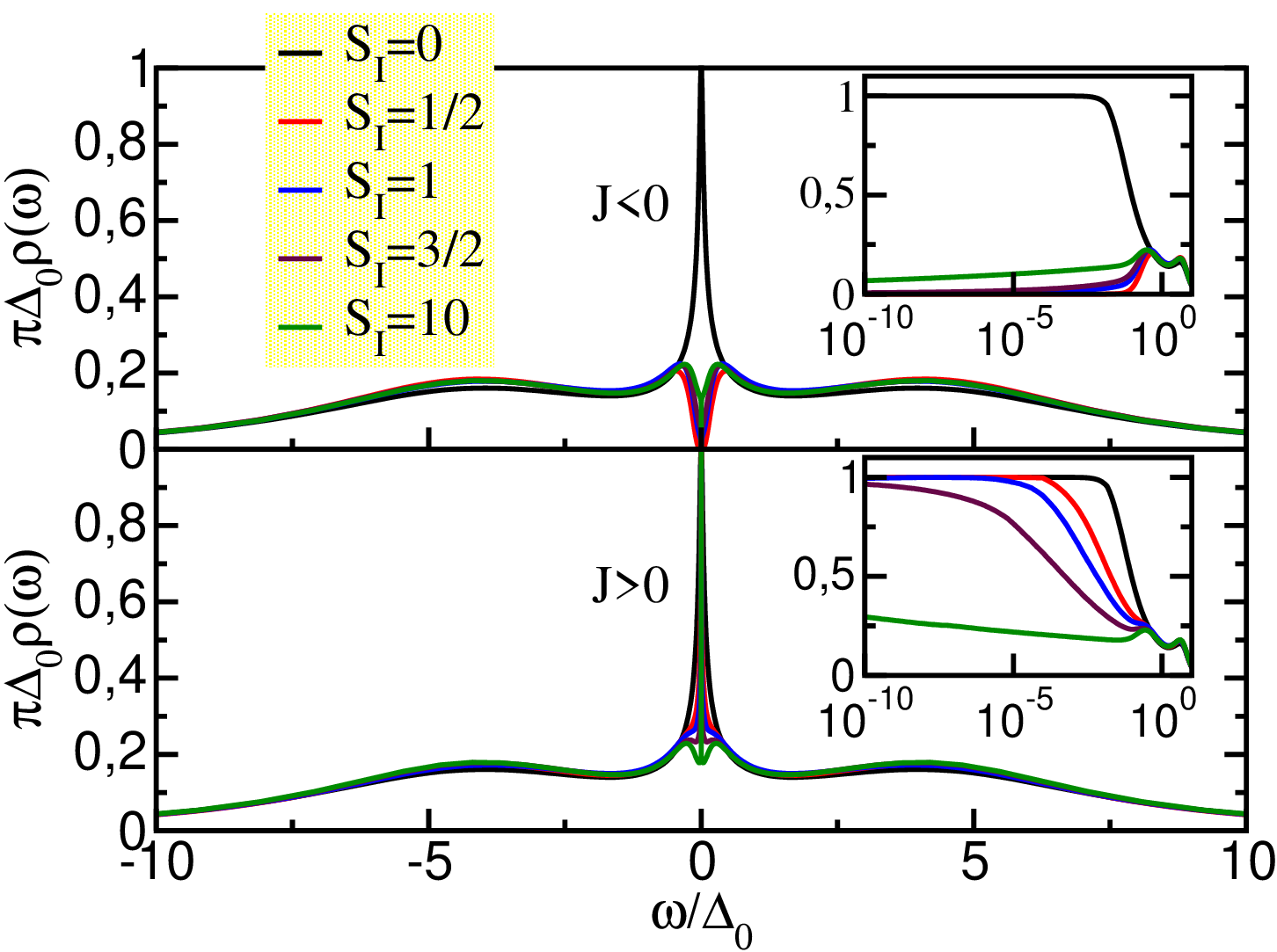}
\end{center}
\caption[]{Local DOS for fixed $\sqrt{S_I(S_I+1)}|J_{\rm K}|/\Delta_0=1/3$ 
as function of $\omega/\Delta_0$ in the intermediate-coupling regime
$U/\Delta_0=10$ for different values of local spin
$S_I$. The upper panel collects the results for antiferromagnetic
coupling, the lower panel those for ferromagnetic.
The insets show
the spectra for $\omega>0$ in a semi-logarithmic plot.\label{fig:dos_inter_c}}
\end{figure}

The differences become even more pronounced if we increase the local
Coulomb repulsion to $U/\Delta_0=10$, which lies in the
intermediate-coupling regime with a Kondo scale $T_{\rm
  K}^{(0)}/\Delta_0\approx0.07$. Since thus $J_{\rm K}>T_{\rm
  K}^{(0)}$ we do not expect two-stage screening here. The results for otherwise same model
parameters are collected in Fig.~\ref{fig:dos_inter_c}.
Note that due to the choice of $J_{\rm K}$ the Hubbard bands do not
move with increasing $S_I$. Furthermore, in all cases $S_I>0$ additional
structures appear at $\omega\approx \pm|J_{\rm K}|$. The low-energy behavior, however, is
markedly different for $J_{\rm K}<0$ (upper panel in
Fig.~\ref{fig:dos_inter_c}) and $J_{\rm K}>0$ (lower panel in
Fig.~\ref{fig:dos_inter_c}). Below $\omega/\Delta_0<0.07\approx T_{\rm
K}^{(0)}$, the former
case always develops a (pseudo-) gap due to the formation of a singlet
between the local degrees of freedom, while the latter tends to
recover a Kondo-resonance for the total spin, again with exponentially decreasing $T_{\rm
K}(S_I)$. Owing to the universal behavior of the conventional Anderson
model for $U/\pi\Delta_o>1$ (``Kondo regime'') we actually expect the
behavior observed here to be generic in this parameter regime. Quite
obviously, while in the weak coupling regime the ``classical limit''
is reached already for moderate values of $S_I$, one has to be very
careful when dealing with the strongly correlated regime.

\section{Summary and conclusion}
In this paper we presented calculations for an extended Anderson
impurity model, where the local charge degrees of freedom in addition
couple to a localized spin. The motivation to study such a model is
based on the observation that in a variety of transition metal or
rare earth compounds the complex local orbital structure can be split
into a localized spin, which can take large values $S\gg1/2$, coupled
via Hund's exchange to a more delocalized set of possibly also
correlated states. A particular example surely is the famous
LaMnO$_3$.

The solution of this model for different regimes of model parameters
was accomplished by using Wilson's NRG, which provides accurate
and reliable results for thermodynamics and dynamics and is able to
resolve arbitrarily small energy-scales that may appear in the
problem. The findings can be summarized as follows: Even for
comparatively large localized spin $S_I=10$, we still observe the
influence of quantum fluctuations on the properties of the impurity
charge degrees of freedom. These effects become more pronounced when
these charge degrees of freedom are correlated themselves as to be
expected for example in LaMnO$_3$. Depending on the ratio $|J_{\rm
  K}|/T_{\rm K}^{(0)}$, where $T_{\rm K}^{(0)}$ is the Kondo
temperature for the model without additional spin, different regimes
can be identified, which in contrast to the classical prediction do
markedly depend on the sign of $J_{\rm K}$.

Thus, for the solution of correlated lattice models with such an
additional spin degree of freedom within DMFT one has to be likely
careful when using the approximation of a classical spin, even when
$S_I$ is comparatively large. Moreover, the expected physics can be
read off our results right away, at least for half-filling. Due to the
reduced Kondo scale for Hund's type or ferrmomagnetic coupling, we
expect a corresponding reduction of a  critical $U$ for a Mott-Hubbard
transition. On the other hand, for antiferromagnetic exchange coupling
the corresponding $U_c=0$ at $T=0$, because the forming of a local
singlet immediately leads to an insulating state.

Quite interesting are also the magnetic properties of the
system. Again, we may anticipate from the impurity calculations that
antiferromagnetism is still the prevailing magnetic order, but in the
vicinity of certain commensurate band fillings we can also expect
ferromagnetic order from a corresponding RKKY exchange. These
inverstigations are currently in progress.

\begin{acknowledgments}
We acknowledge useful conversations with
M.\ Vojta,
A.\ Lichtenstein,
R.\ Bulla,
F.\ Anders
and
D.\ Vollhardt.
This work was supported by the DFG through the collaborative research
center SFB 602. Computer support was provided through the Gesellschaft
f\"ur wissenschaftliche Datenverarbeitung in G\"ottingen and
the Norddeutsche Verbund f\"ur Hoch- und H\"ochstleistungsrechnen.

\end{acknowledgments}

%\bibliographystyle{apsrev}
%\bibliography{../Bibliography/bibfile}

\end{document}